\documentclass[aps,prl,reprint,showpacs,floatfix,longbibliography,citeautoscript,superscriptaddress]{revtex4-1}
\usepackage{graphicx}
\usepackage{bm,amsmath,amssymb,mathrsfs}
\usepackage[colorlinks=true, linkcolor=blue, citecolor=blue, urlcolor=blue, linktoc=page, bookmarks=false, pdfstartview={FitH}, pdfborder={0 0 0.0 [3 3]}]{hyperref} 
\usepackage[usenames]{color}
\usepackage{longtable}
\usepackage{dcolumn}
\usepackage{subfigure}
\usepackage{units}
\usepackage[final]{pdfpages}
\hypersetup{pdfborder=0 0 0,colorlinks=true,citecolor=blue,linkcolor=blue}

\DeclareMathOperator{\sgn}{sgn}

\begin{document}

\title{$\mathbb{Z}_2$ invariance of Germanene on MoS$_2$ from first principles}

\author{Taher Amlaki}
\affiliation{Faculty of Science and Technology amd MESA$^+$ Institute for Nanotechnology,
University of Twente, P.O. Box 217, 7500 AE Enschede, The Netherlands}

\author{Menno Bokdam}
\affiliation{Faculty of Physics, University of Vienna, Computational
Materials Physics, Sensengasse 8/12, 1090 Vienna, Austria}

\author{Paul J. Kelly}
\affiliation{Faculty of Science and Technology amd MESA$^+$ Institute for Nanotechnology,
University of Twente, P.O. Box 217, 7500 AE Enschede, The Netherlands}


\begin{abstract}
We present a low energy Hamiltonian generalized to describe how the energy bands of germanene ($\rm \overline{Ge}$) are modified by interaction with a substrate or a capping layer. The parameters that enter the Hamiltonian are determined from first-principles relativistic calculations for $\rm \overline{Ge}|$MoS$_2$ bilayers and MoS$_2|\rm \overline{Ge} |$MoS$_2$ trilayers and are used to determine the topological nature of the system. For the lowest energy, buckled germanene structure, the  gap depends strongly on how germanene is oriented with respect to the MoS$_2$ layer(s). Topologically non-trivial gaps for bilayers and trilayers can be almost as large as for a free-standing germanene layer.     
\end{abstract}

\maketitle

{\color{red}\it Introduction.}---Insulators can be categorized by topological invariants that are not continuous; when these have to change, interesting physics occurs. The first group of these invariants was found to describe the quantum Hall effect for electrons confined in strong magnetic fields \cite{Zhang:nat05, Bolotin:ssc08, Hasan:rmp10}. A new class of ``topological'' insulators (TI) was proposed for systems with time-reversal symmetry where the invariant can have two values \cite{Kane:prl05a, Kane:prl05b} and topologically non-trivial systems are called $\mathbb{Z}_2$ TIs \cite{Kane:prl05a, Kane:prl05b, Fu:prl07, Fu:prb07, Bernevig:sc06, Koenig:sc07}. In the two dimensional (2D) graphene ($\rm \overline{C}$) originally studied by Kane and Mele \cite{Kane:prl05a, Kane:prl05b}, spin-orbit coupling (SOC) leads to the opening of a gap at the Dirac point giving rise to the possibility of topologically protected spin-polarized edge states. The intrinsic SOC of carbon is, however, very small resulting in gaps of less than 50~$\mu$eV (0.6~K) \cite{Min:prb06, *Huertas:prb06, *Yao:prb07, *Boettger:prb07, *Gmitra:prb09}. 
Two approaches have been taken to resolve this issue. One is to induce a larger spin-orbit coupling in graphene by placing it in contact with layered materials that contain heavy elements with large intrinsic SOC \cite{Kou:nanol13, *Kaloni:apl14, *Kou:acsn14, *Kou:carb15, *Kou:acsami15, Wang:natc15, Gmitra:prb15}. 
The other is to begin with a 2D group IV material with a larger intrinsic SOC \cite{[For recent reviews with additional references{,} see ]Balendhran:sma15, *Acun:jpcm15}. Motivated by recent success in growing germanene ($\rm \overline{Ge}$) on MoS$_2$ \cite{ZhangL:prl16}, this paper is concerned with the latter.

The structures and stability of free-standing group IV layers have already been studied theoretically. Both silicene ($\rm \overline{Si}$) and germanene ``buckle'' with the two sublattices moving in opposite directions out of the original plane but maintaining inversion symmetry \cite{Takeda:prb94, GuzmanVerri:prb07, Cahangirov:prl09}; stanene ($\rm \overline{Sn}$) forms a different dumbell structure \cite{Tang:prb14}. The unsupported layers are predicted to be TIs \cite{LiuCC:prl11, Tang:prb14}. Experimental efforts have so far focussed on growing silicene \cite{LewYanVoon:mrsb14} and germanene \cite{Acun:jpcm15} on metallic substrates where the intrinsic transport properties cannot be studied. Eventually these layered structures must be transferred to or grown on a nonconducting substrate. It is then essential to know if the TI character survives the interaction with the substrate. However, the complexity of these systems has made calculation of the topological invariant impossible until now.

We focus on the recently grown $\rm \overline{Ge}|$MoS$_2$ system \cite{ZhangL:prl16}. A free-standing, planar germanene layer has a SOC induced gap of 4~meV. Buckling breaks the reflection symmetry, mixes the $p_z$ with the $\{s,p_x,p_y\}$ orbitals and increases the SOC gap to 24~meV \cite{LiuCC:prl11}. It leads to one Ge sublattice interacting more strongly with a substrate than the other, breaking the sublattice symmetry and opening a gap as large as $\sim 40$~meV without SOC; with SOC included, Rashba SOC is induced by the breaking of reflection (and inversion) symmetry. To investigate whether or not the gapped asymmetric  bilayer is a TI, we generalize Kane and Mele's model to describe the interaction with a substrate. 
We use first-principles calculations to determine equilibrium geometries,  to evaluate the parameters in the model Hamiltonian from the first-principles electronic structures and to calculate phase diagrams. We will identify the orientation of germanene on the substrate as the most critical factor in determining the size and topological nature of the band gap. The SOC induced  band gap of free-standing $\rm \overline{Ge}$ can be almost completely restored in an MoS$_2|{\rm \overline{Ge}}|$MoS$_2$ trilayer where the sandwich structure should stabilize and protect the $\rm \overline{Ge}$ layer from the environment.

{\color{red}\it Phenomenological model: asymmetric bilayer.}---We begin by constructing a low energy Hamiltonian for graphene interacting (weakly) with a semiconducting substrate (S) by downfolding a tight-binding (TB) Hamiltonian for the same system. Taking ${\bm \sigma}$ and ${\bf s}$ to be vectors of Pauli matrices where ${\bm \sigma}$ represents the $A(B)$ sublattices of graphene and ${\bf s}$ represents spin, then the result for an asymmetric (AS) $\rm \overline{C}|$S bilayer is
\begin{equation}
H_{\bf K}^{\rm AS}({\bf q}) = \hbar v_F{\bf q}.{\bm \sigma} +\lambda_m \sigma_z + \frac{\lambda_R}{2}({\bm \sigma} \times {\bf s})_z + \lambda_{\rm so} \sigma_z s_z + \lambda _Bs_z 
\label{eqq1}
\end{equation}
where ${\bf q}$ is the wave vector relative to the ${\bf K}$ point, ${\bf q}={\bf k}-{\bf K}$. $\lambda_m$ is a ``mass'' term that describes the breaking of the sublattice symmetry by the interaction with the substrate. $\lambda_R$ is a Rashba SOC term that results from the breaking of reflection symmetry in the direction perpendicular to the germanene layer. $\lambda_{\rm so}$ is Kane and Mele's spin-orbit term \cite{Kane:prl05a} that contains the intrinsic ``atomic'' SOC term of monolayer germanene plus $\lambda_{\rm so}^{\rm (ind)}$, the SOC induced by the substrate. $\lambda_B$ corresponds to a ``pseudomagnetic'' term which is odd under inversion symmetry and changes sign at the ${\bf K}'$ point and therefore does not break time-reversal symmetry. 

\begin{figure}[t]
\includegraphics[width=8.0cm]{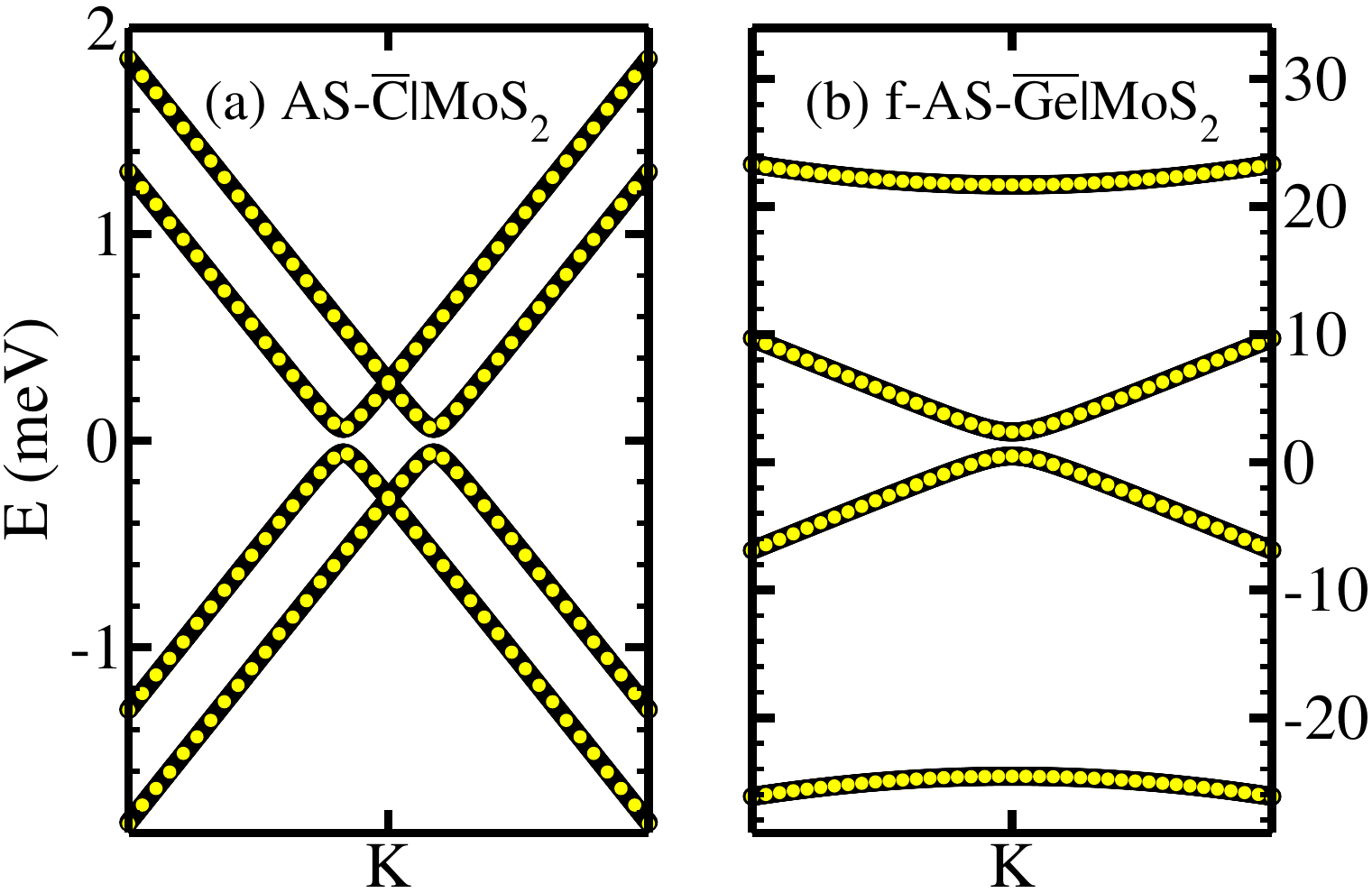}
\caption{
Band structures of (a) AS p-$\rm \overline{C}|$MoS$_2$ and (b) AS b-$\rm \overline{Ge}|$MoS$_2$ bilayers close to the ${\bf K}$ point. The yellow dots are the results of first principles calculations, the black lines result from the model \eqref{eqq1} with parameters from \eqref{eqq5}.
}
\label{fig1}
\end{figure}

The eigenvalues of \eqref{eqq1} at the ${\bf K}$ point are
\begin{subequations}
\begin{eqnarray}
 \varepsilon_{4(3)}=&& \lambda_{\rm so} \pm( \lambda_B + \lambda_m) \\
 \varepsilon_{2(1)}=&-&\lambda_{\rm so} \pm \sqrt{(\lambda_B-\lambda_m)^2+\lambda_R^2}\,
\end{eqnarray}
\label{eqq2}
\end{subequations}
By comparing these eigenvalues and the corresponding eigenvectors with those calculated from first-principles, we can determine the parameters in \eqref{eqq1} with which the band structure about the Dirac point can be described. The projection of wavefunctions onto specific atoms is not unique. However, the spin space is complete to very good accuracy and we use the expectation values for the $z$ component of spin
\begin{equation}
\langle s_z \rangle_{n{\bf K}}=\frac{1}{\Omega }\int_{\Omega} \left(
 \left|\psi_{n{\bf K}}^{\uparrow }  ({\bf r})\right|^2
-\left|\psi_{n{\bf K}}^{\downarrow }({\bf r})\right|^2 \right)d^2{\bf r}
\label{eqq3}
\end{equation}
for the four bands at the Dirac point where the integral should be taken over the supercell with area $\Omega$. 
Applying \eqref{eqq3} to first principles results to be presented below shows that $\langle s_z \rangle_{\bf K}$=($s$,-$s$,-1,1) for the four bands at the Dirac point; here $s$ is a positive number smaller than one. Solving for the parameters in \eqref{eqq1} results in 
\begin{subequations}
\begin{eqnarray}
\lambda_m       =&&\,[(\varepsilon_4-\varepsilon_3)+s(\varepsilon_2-\varepsilon _1)]/4 \\
\lambda_R       =&&\,\pm (\varepsilon_2-\varepsilon_1)\sqrt{1-s^2} /2 \\
\lambda_{\rm so}=&&\,[(\varepsilon_4+\varepsilon_3)-(\varepsilon_2+\varepsilon _1)]/4 \\
\lambda_B       =&&\,[(\varepsilon_4-\varepsilon_3)-s(\varepsilon_2-\varepsilon _1)]/4
\end{eqnarray}
\label{eqq5}
\end{subequations}
When buckling is included, the TB Hamiltonian cannot be exactly downfolded. However, it does not introduce any qualitatively new symmetries and \eqref{eqq1} describes the band dispersion about the Dirac point equally well for planar $\rm \overline{C}|$MoS$_2$ and buckled $\rm \overline{Ge}|$MoS$_2$ as seen in Fig.~\ref{fig1}. 

{\color{red}\it First-principles calculations.}---We use density functional theory (DFT) to calculate ground state energies and optimized geometries with a projector augmented wave (PAW) basis \cite{Blochl:prb94b,Kresse:prb99} as implemented in \textsc{vasp} \cite{Kresse:prb93, Kresse:prb96} for $\rm \overline{Ge}|MoS_2$ bilayers and MoS$_2| \rm \overline{Ge}|MoS_2$ trilayers \cite{SM}. 
We first determine equilibrium geometries for individual monolayers of $\rm \overline{Ge}$ and MoS$_2$. For germanene, both planar (p-$\rm \overline{Ge}$) and buckled (b-$\rm \overline{Ge}$) structures are studied. For relaxed b-$\rm \overline{Ge}$ the sublattice planes are separated by $c=0.71$\AA. The calculated in-plane lattice constants are 4.05, 4.05 and 3.16 \AA\ for p-$\rm \overline{Ge}$, b-$\rm \overline{Ge}$ and MoS$_2$, respectively. We identify lattice vectors in both materials with an acceptable length mismatch and then rotate the two lattices through an angle $\theta$ to make them coincide; this defines a ``supercell''. 

Because of the weak interaction between germanene and MoS$_2$ a strong preference for a particular alignment of the two lattices is not expected and this is borne out by the weak binding energy we find for the relaxed structures. We accommodate the small residual lattice mismatch in the MoS$_2$ layer and reoptimize its structure. The $\rm \overline{Ge}$ and MoS$_2$ layers are allowed to bond in two stages, first only changing the height of the b-$\rm \overline{Ge}$ above MoS$_2$ (h-AS structure) and then without constraint (f-AS structure). For a supercell, the average buckling is calculated as $c =\sqrt{\sum_{N_{\rm Ge}} c_i^2/N_{\bf Ge}}$ and is given together with other relevant parameters in Table~\ref{tb1} for the smallest ``reasonable sized'' supercell containing 89 atoms with $\theta=24.8^{\circ}$ and an acceptable lattice mismatch of 0.7\%. For the h-AS bilayer, the separation of the bottom germanene plane from the upper sulphur layer is 3.11~\AA.


\begin{table}[b]
\vspace{-1em} 
\caption{$E_b$ is the binding energy in meV per $\rm \overline{Ge}$ unit cell. The dimensionless spin parameter $s$ is defined in the text. $\Delta_{\bf K}$ is the gap calculated at the ${\bf K}$ point in meV. The Hamiltonian parameters defined in equations \eqref{eqq2} and \eqref{eqq5} are given in meV for free-standing planar and buckled $\rm \overline{Ge}$ layers, for AS $\rm \overline{Ge}|$MoS$_2$ bilayers and for IS MoS$_2|{\rm \overline{Ge}}|$MoS$_2$ trilayers. $c$ is the separation between the two Ge planes in \AA and $v_F\approx 4\times 10^5m/s$. For f-AS $\rm \overline{C}$, shown for comparison, the minimum gap is not at ${\bf K}$.
}
\begin{ruledtabular}
\begin{tabular}{lrrrrrrrrrrr}
                      &$E_b$ 
                            & $s$  & $\Delta_{\bf K}$
                                           & $\lambda_m$
                                                   & $\lambda_{\rm {so}}$
                                                           & $\lambda _R$
                                                                  & $\lambda_B$
                                                                          & $c$(\AA) \\
\hline
p-$\rm \overline{Ge}$ &  -- &   -- &  4.21 &    -- &  2.11 &   -- &    -- & 0.00\\
b-$\rm \overline{Ge}$ &  -- &   -- & 25.78 &    -- & 12.89 &   -- &    -- & 0.71\\
h-AS                  & 328 & 0.83 &  5.55 &  7.95 & 11.60 & 5.72 & -0.56 & 0.71\\
f-AS                  & 332 & 0.87 &  1.88 & 10.28 & 12.04 & 6.18 & -0.62 & 0.73\\
\hline 
f-AS ($\rm \overline{C}$)
                      &  45 & 0.91 &  0.55 & -0.08 &  0.00 & 0.12 & -0.27 & 0.00\\
\hline 
h-IS                  & 671 &   -- & 21.21 &   --  & 10.61 &   -- &    -- & 0.71\\
f-IS                  & 680 &   -- & 22.71 &   --  & 11.36 &   -- &    -- & 0.75\\
\end{tabular}
\end{ruledtabular}
\label{tb1}
\end{table}

{\color{red}\it Results: AS bilayers.}---The band structures of p-$\rm \overline{C}|$MoS$_2$ and b-$\rm \overline{Ge}|$MoS$_2$ bilayers close to the Dirac point are compared in Fig.~\ref{fig1}. On this small energy scale, the shape of the bands is quite different because $\lambda_B$ is dominant for graphene while for germanene $\lambda_{\rm so}$, $\lambda_m$, and $\lambda_{R\rm }$ are much larger. It is clear from the figure that the phenomenological model (black lines) describes the low energy first-principles bands (yellow dots) close to the ${\bf K}$ point very accurately for different regimes. For AS b-$\rm \overline{Ge}|$MoS$_2$ the gap decreases from 5.6 meV for the height optimized structure (h-AS) to 1.9 meV for the fully unconstrained structure (f-AS); see Table~\ref{tb1}. $\lambda_m$ is there seen to increase faster than $\lambda_{\rm so}$ because the average buckling increases slightly from 0.71 to 0.73\AA\ so the gap decreases. Another important point is that $\lambda_{\rm so}^{\rm (ind)}$ is negative. Calculating $\lambda_{\rm so}^{\rm (ind)}=\lambda_{\rm so}^{\rm h-AS}-\lambda_{\rm so}^{\rm b-\overline{Ge}}$ with parameters from Table~\ref{tb1} yields $\lambda_{\rm so}^{\rm(ind)}=11.60-12.89=-1.29$ meV and therefore interaction with the MoS$_2$ layer reduces the intrinsic SOC induced gap of germanene. The mass and Rashba terms are larger than the induced SO term and both $\lambda_m$ and $\lambda_R$ increase faster than $\lambda_{\rm so}^{\rm (ind)}$ if the interaction between germanene and MoS$_2$ increases. Applying pressure to AS $\overline {\rm Ge}|$MoS$_2$ reduces the gap until $\lambda_{\rm so}=\frac{1}{2}\big(\lambda_m + \lambda_B + \sqrt{\left(\lambda_m-\lambda_B\right){}^2 + \lambda_R^2}\big)$ when it vanishes. After that, the band gap grows again but the topological nature of the bands changes. Applying pressure to AS $\overline{\text {Ge}}|$MoS$_2$ will therefore not result in a TI with a larger band gap.

To determine the $\mathbb{Z}_2$ topological invariant $\nu$ for the AS system, we analyse the phase space corresponding to \eqref{eqq1} with the parameter values from Table~\ref{tb1}. $\nu$ is related to the integral of the Berry curvature $B({\bf q})$ over the effective Brillouin zone (EBZ) and the Berry potential over its boundary \cite{Essin:prb07}. In our four band model the full Brillouin zone is ${\bf K} \oplus {\bf K}'$, the EBZ contains only ${\bf K}$ and therefore 
\begin{equation}
\nu = \left\{1 + \frac{1}{2\pi} \int \Big[ B_1({\bf q})+B_2({\bf q}) \Big] d{\bf q} \right\} \bmod 2
\label{eqq8}
\end{equation}
\noindent
where $B_i({\bf q})$ is the Berry curvature of the $i^{\rm th}$ band and unity in the curly brackets is the contribution of the boundary. Since it is a topological invariant $\nu$ will not change unless the band gap vanishes so TI and NI regions should be separated by zero-gap lines. According to \cite{Kane:prl05b}, the system will be a TI if the $\lambda_{\rm so}$ term is dominant whereas if $\lambda_m$ is dominant, the system will be a NI. Any point in the phase space that can be connected to any of the $\lambda_{\rm so}$ dominated points without closing the gap is TI.

\begin{figure}[t]
\includegraphics[width=7.0cm]{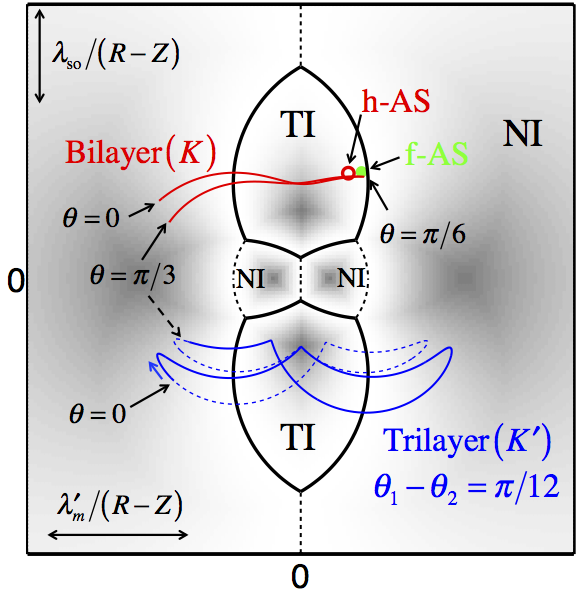}
\caption{Stereographic projection of the phase space of the Hamiltonian \eqref{eqq1}. Black lines represent boundaries between regions where the gap vanishes; phases on either side of the dashed black lines are the same. The scaling of the $\lambda_{\rm so}$ and $\lambda_m$ variables with $R-Z$ is explained in the text. When germanene is rotated with respect to MoS$_2$, a trajectory is traced out in parameter space which is shown in red for a $\rm \overline{Ge}|$MoS$_2$ bilayer and in blue for a MoS$_2|\rm \overline{Ge} |$MoS$_2$ trilayer where the two MoS$_2$ layers are rotated with respect to one another by $\theta_1-\theta_2=15^{\circ}$. }
\label{fig2}
\end{figure}

The general phase space for the Hamiltonian \eqref{eqq1} is four dimensional. Scaling all the parameters will result in scaling of all the eigenvalues so we only need to study the surface of a sphere ($S^3$) of radius $R$ ($R^2=tr\left.H^2\right/4$) in this four dimensional space. Since there are only three independent eigenvalues, we construct a map $\phi :S^3\rightarrow S^2$ where 
$X \equiv \lambda'_m=(\lambda_m+\lambda_B)/\sqrt{2}$, 
$Y=\lambda _{\text{so}}$, 
$Z=\sqrt{(\lambda_R^2+(\lambda_m-\lambda_B)^2)/2}$ and 
$X^2+Y^2+Z^2=R^2$. Adding a term $-Z$ to symmetrize $\phi$, the eigenvalues at ${\bf K}$ will be $\varepsilon_{4(3)}=Y\pm \sqrt{2}X$ and $\varepsilon_{2(1)}=-Y\pm \sqrt{2}Z$.
The final step is a conformal map (stereographic projection) $sp: S^2\rightarrow \mathbb{R}^2$ which results in Fig.~\ref{fig2} ($sp(\phi ):S^3\rightarrow \mathbb{R}^2$). 
As long as $|\lambda_B|\leq \sqrt{\lambda_m^2+\lambda_{\rm so}^2}$ - our first-principles calculations will show that this condition is satisfied - the gap remains at the ${\bf K}$ point and will be given by this map. 
The figure show that for $\theta=24.8^{\circ}$ (open red dot), AS b-$\rm \overline{Ge}|$MoS$_2$ is a topological insulator  - just. Relaxing the germanene layer fully on MoS$_2$ does not change the $\mathbb{Z}_2$ invariant though the reduced gap means that it is less stable (green dot). 

\begin{figure}[t]
\includegraphics[width=0.90\columnwidth]{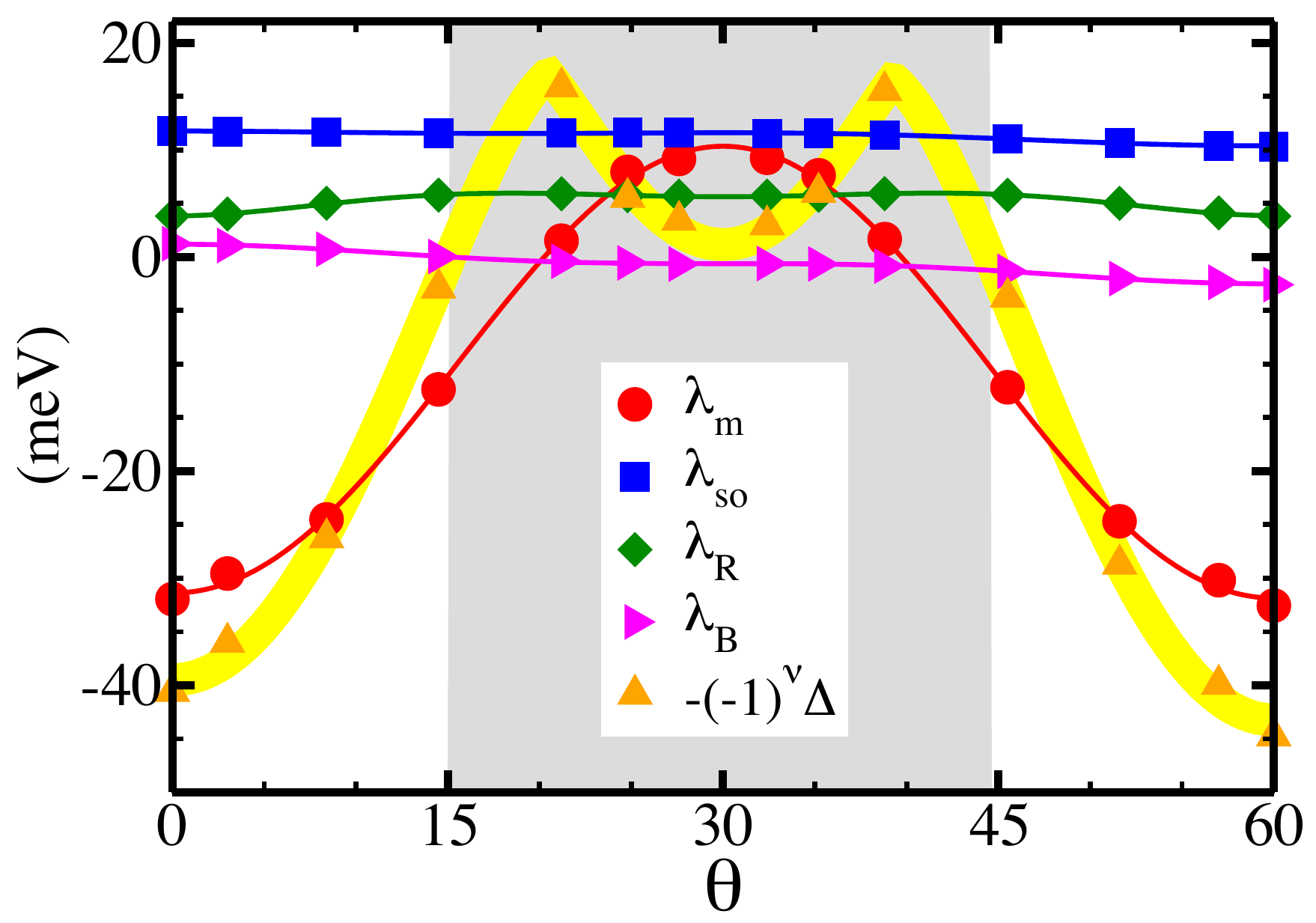}
\caption{$\lambda$ parameters as function of the angle $\theta$ for a fixed height of germanene above MoS$_2$ that minimizes the energy for $\theta=24.8^{\circ}$ for b-$\rm \overline{Ge}|MoS_2$. The dashed lines are fits to expressions with appropriate angle symmetries. Details of the calculations and the parameters extracted for both planar and buckled $\rm \overline{Ge}|MoS_2$ can be found in \cite{SM}.}
\label{fig3}
\end{figure}

\begin{figure}[t]
\includegraphics[width=0.80\columnwidth]{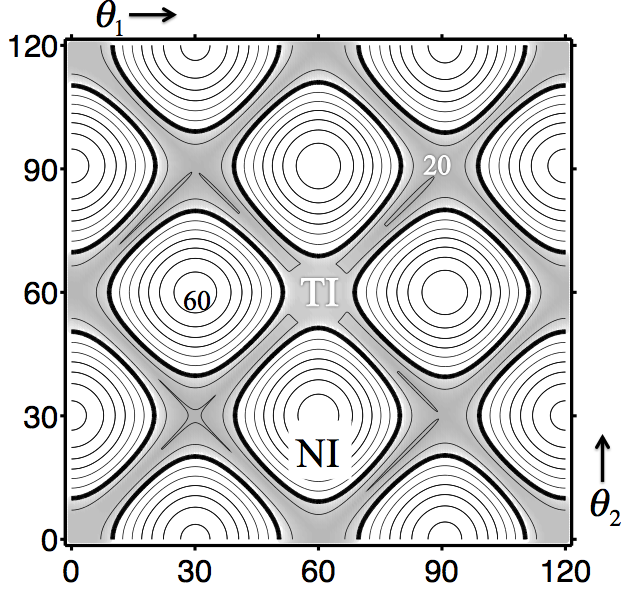}
\caption{Dependence of the band gap on the angles $\theta_1$ and $\theta_2$ that a germanene layer makes with two MoS$_2$ layers in a MoS$_2|\rm \overline{Ge} |$MoS$_2$ trilayer with threefold rotation symmetry. The unshaded region is NI, the shaded region TI.}
\label{fig4}
\end{figure}

For planar germanene (or graphene \cite{Amlaki:16}), the $\lambda$ parameters depend only weakly on the orientation with respect to the MoS$_2$ substrate \cite{SM}. Buckling brings one germanene sublattice into closer contact with the substrate than the other and this leads to a non-vanishing mass term $\lambda_m$. When germanene is displaced parallel to the substrate $\lambda_m$ varies very weakly \cite{SM} but when it is rotated through some angle $\theta$ it varies strongly as shown in Fig.~\ref{fig3} (red dots and curve). This gives rise to a much more complex dependence of the gap on the germanene orientation, $\Delta_{\bf K}(\theta)$ (yellow dots and curve). The angle dependence of the other parameters is seen to be much smaller. The shaded part of Fig.~\ref{fig4} is TI and for AS b-$\rm \overline{Ge}|MoS_2$ bilayers a sizeable gap of more than 15 meV is predicted for angles $\theta \sim 20^{\circ}$ and $\theta \sim 40^{\circ}$. In the phase diagram Fig.~\ref{fig2}, the full angle dependence is shown  as a red line. 

{\color{red} MoS$_2|\rm \overline{Ge} |$MoS$_2$ {\it trilayers}.}---In an experiment it will be necessary to protect the germanene layer. A second, capping layer of MoS$_2$ will most likely be at some arbitrary angle $\theta_2$ to germanene, itself at an angle $\theta_1$ to the substrate MoS$_2$ layer, making it important to know how the gap will depend on $\theta_1$ and $\theta_2$. The large separation of the two MoS$_2$ layers suggests that the direct interaction can be neglected in our TB derivation, leading to the prediction that the effect of the two MoS$_2$ layers will be additive in terms of the parameters in \eqref{eqq1}. This is confirmed by explicit calculation for trilayers with $(\theta_1,\theta_2)=(24.8^{\circ},24.8^{\circ})$, $(24.8^{\circ},3.0^{\circ})$, and $(3.0^{\circ},3.0^{\circ})$ \cite{SM}. The band gap is shown as a function of  $\theta_1$ and $\theta_2$ in Fig.~\ref{fig4}. The NI gap can be in excess of 60 meV when the $\lambda_m$ contributions do not cancel. The TI gap is largest ($> 20$~meV) when they cancel exactly for $\theta_1 \pm \theta_2= n\pi/3$ for integer $n$.

{\color{red}\it Inversion symmetric trilayer.}---
The term containing $\lambda_R$ in Eq.~\ref{eqq1} is odd under inversion. For an MoS$_2|\rm \overline{Ge}|$MoS$_2$ trilayer constructed to have inversion symmetry (IS) the average of $\lambda_R$ over a supercell is zero so this term is absent. The mass term $\lambda_m$ and pseudomagnetic term $\lambda_{\rm B}$ also vanish because they are odd under inversion and \eqref{eqq1} simplifies to $H_{\bf K}^{\rm IS}({\bf q})=\hbar v_F{\bf q}.{\bm \sigma} + \lambda_{\rm so} \sigma_z s_z$.
This equation satisfies the requirement of Kramers degeneracy that all bands should be doubly degenerate and predicts that the gap will vanish only if $\lambda_{\rm so}$ is zero. In this case $\left\langle s_z\right\rangle$ is not uniquely defined  because degenerate bands have complementary spin textures.

Using the effective Hamiltonian parameters calculated for the AS b-$\rm \overline{Ge}|$MoS$_2$ bilayer with $\theta=24.8^{\circ}$, we can estimate the band gaps at the ${\bf K}$ point for the IS MoS$_2|{\rm \overline{Ge}}|$MoS$_2$ trilayer. 
For the h-AS system $\lambda_{\rm so}^{\rm (ind)}$ was found to be $-1.29$~meV. For the h-IS configuration, we predict $\lambda_{\rm so}^{\rm (IS)}=\lambda_{\rm so}^{\rm (Ge)}+2\lambda_{\rm so}^{\rm (ind)}=12.89-2\times 1.29=10.31$. An explicit first-principles calculation yields a value of $\lambda_{\rm so}^{\rm (IS)}=10.61$~meV. The close agreement between the predicted and calculated values indicates that the model is consistent \cite{SM}.

For IS systems we can use the formula given by Fu and Kane \cite{Fu:prb07} to determine the TI $\nu$ explicitly from first principles calculations,
\begin{equation}
(-1)^{\nu }=\underset{i=1}{\overset{4}{\Pi }} \,\, \underset{m=1}{\overset{N}{\Pi }}\xi _{2m}\left(\Gamma _i\right)
\end{equation}
where the first multiplication is over all the time-reversal fixed points $\Gamma_i$ and the second multiplication is over bands with even band number at the $\Gamma _i$; $\xi _{2m}$ is the parity eigenvalue of bands $2m-1$ and $2m$. For our inversion symmetric systems, we explicitly calculated the $\mathbb{Z}_2$ invariant and found them all to be topological insulators with band gaps of about 23 meV generated by SO interactions confirming the phase space assignments. 

{\color{red}\it Conclusion.}--- We use a comprehensive phenomenological model to describe spin-orbit interactions for $\rm \overline{Ge}|$MoS$_2$ bilayers and  MoS$_2|{\rm \overline{Ge}}|$MoS$_2$ trilayers. We determine the parameters entering this model from the eigenvalues and spin expectation values at the ${\bf K}$ point. The model describes the low energy band structure of germanene very accurately and provides insight into the different interactions involved. For a $\rm \overline{Ge}|$MoS$_2$ bilayer the band gap of germanene is dominated by the mass term $\lambda_m$ that depends strongly on how germanene is oriented on the MoS$_2$ substrate. A maximum non-trivial TI gap of $\sim 15$~meV is predicted for angles of $20^{\circ}$ and $40^{\circ}$. By sandwiching $\rm \overline{Ge}$ between MoS$_2$ layers, the large 24 (26) meV intrinsic SOC gap reported \cite{LiuCC:prl11} (we find) for free-standing germanene can be almost fully recovered but requires being able to control the orientation of germanene with respect to both MoS$_2$ layers. Exploratory many-body corrections to these single particle gaps indicate that they may be enhanced by an order of magnitude.


\begin{acknowledgments}We are grateful to Harold Zandvliet for communicating the results of \cite{ZhangL:prl16} before publication. T.A. acknowledges fruitful discussions with Mojtaba Farmanbar. This work is part of the research program of the Foundation for Fundamental Research on Matter (FOM), which is part of the Netherlands Organisation for Scientific Research (NWO).  The use of supercomputer facilities was sponsored by the Physical Sciences division of NWO (NWO-EW).
\end{acknowledgments}

\pagebreak
\begin{widetext}
\clearpage
\begin{center}
\textbf{\large Supplementary Material for ``$\mathbb{Z}_2$ invariance of Germanene on MoS$_2$ from first principles''}

\vspace{3mm}
Taher Amlaki,$^{1,\ast}$ Menno Bokdam,$^2$, and Paul J. Kelly$^1$

\vspace{2mm}
$^1${\small\it Faculty of Science and Technology and MESA$^+$ Institute for Nanotechnology, \\University of Twente, P.O. Box 217, 7500 AE Enschede, The Netherlands}

$^2${\small\it Faculty of Physics, University of Vienna, Computational
Materials Physics, Sensengasse 8/12, 1090 Vienna, Austria}

\end{center}
\end{widetext}

\setcounter{equation}{0}
\setcounter{figure}{0}
\setcounter{table}{0}
\setcounter{page}{1}
\makeatletter
\renewcommand{\theequation}{S\arabic{equation}}
\renewcommand{\thefigure}{S\arabic{figure}}
\renewcommand{\thetable}{S\Roman{figure}}
\renewcommand{\bibnumfmt}[1]{[S#1]}
\renewcommand{\citenumfont}[1]{S#1}

\subsection{A. Computational details.}

Density functional theory (DFT) was used to calculate ground state energies and optimized geometries for $\rm \overline{Ge}|MoS_2$ bilayers and MoS$_2| \rm \overline{Ge}|MoS_2$ trilayers with a projector augmented wave (PAW) basis \cite{Blochl:prb94bs,Kresse:prb99s} as implemented in the \textsc{vasp} \cite{Kresse:prb93s, Kresse:prb96s} code. The bilayers and trilayers were repeated periodically and separated from their images by a 15\,\AA\ thick vacuum region. A dipole correction was applied to avoid spurious interactions between the  periodic images \cite{Neugebauer:prb92s}. The plane wave kinetic energy cutoff was set at 600 eV. We used a dense $42\times 42$ \textbf{k}-point grid to sample the germanene $1\times 1$ Brillouin zone (BZ), and a comparable density for supercells. For BZ integrations we used the tetrahedron scheme \cite{Blochl:prb94as}. The electronic self-consistency criterion was set to $10^{-7}$ eV. Bilayers and trilayers were relaxed until the total energy was converged to within $10^{-7}$ eV. The high level of precision is necessary to study band gaps of order meV reproducibly. As a compromise between the LDA density functional (DF) that tends to overbind, and GGA that underbinds van der Waals (vdW) structures, we used the optB88-vdW-DF \cite{Dion:prl04s, Thonhauser:prb07s, Klimes:prb11s}.

The SO splitting of the $p$ valence states of free C, Si and Ge atoms, 8.7, 32.5 and 190.2 meV, respectively has a negligible effect on the equilbrium structures of the corresponding monolayers. Compared to C and Si, the increase in the atomic SOC for Ge is reflected in a greatly increased SO splitting for the planar monolayer which in turn is greatly enhanced by buckling; the ${\bf K}$-point SO splitting of p-$\rm \overline{Ge}$ is 4.2~meV, that of b-$\rm \overline{Ge}$ is 25.8~meV. For a free-standing $\rm \overline{Ge}$ layer, buckling preserves inversion symmetry so the bands at the Dirac point are linear in ${\bf q}$ in the absence of SOC. The parameters extracted from the first principles calculations and Eqs.~(2), (3) and (4) are given in Table~I.

\begin{figure}[b]
\includegraphics[width=8.0cm]{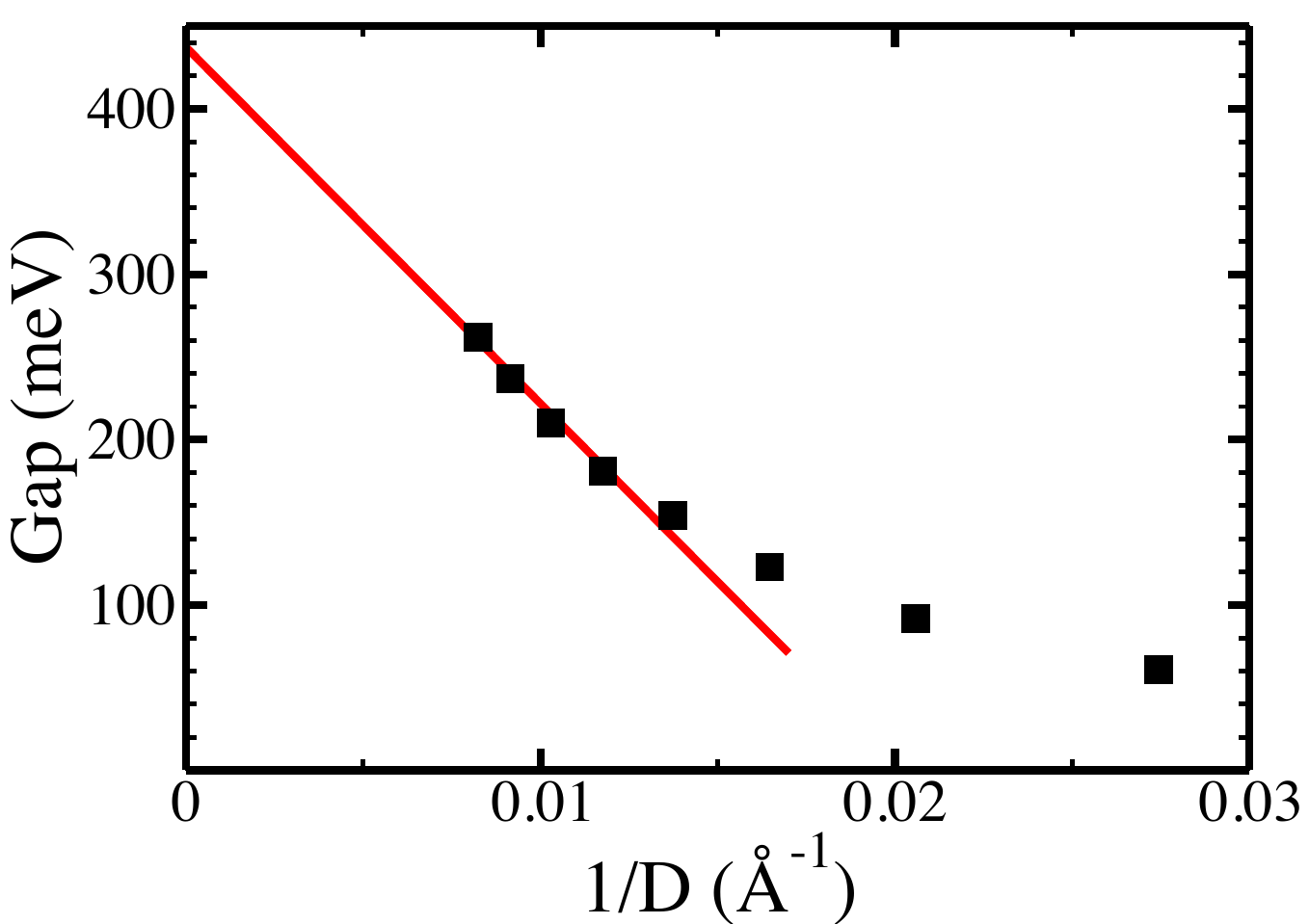}
\caption{Quasiparticle gaps opened at the Dirac point as a function of the inverse cell size. The solid red line interpolates the four largest cell sizes to infinite layer separation.
}
\label{GWgap}
\end{figure}

Many-body effects are studied within the \textit{GW} approximation \cite{Hedin:pr65s} starting with LDA Kohn-Sham (KS) orbitals \cite{Hybertsen:prb86s} for free-standing germanene. We use the \textit{GW} implementation in \textsc{vasp} \cite{Shishkin:prb06s}, with 320 bands and 128 points on the frequency grid. Interactions between periodic images in the $z$ direction lead to a dependence of the \textit{GW} band gap on the cell size $D$. By linearly extrapolating the gaps as a function of the inverse cell size to infinite separation, we can estimate the GW gap for an isolated monolayer \cite{Berseneva:prb13s}. Figure~\ref{GWgap} shows the resulting quasiparticle gaps as a function of  $1/D$. The band gap obtained by extrapolation to $D \rightarrow \infty$ is more than 400 meV, a dramatic increase on the 26 meV LDA value (not shown). 

To very good accuracy the spin space for the four germanene derived bands at the ${\bf K}$ point is complete. To calculate the expectation values for the $z$ component of spin
\vspace{-1em}
\begin{equation}
\langle s_z \rangle_{n{\bf K}}=\frac{1}{\Omega }\int_{\Omega} \left(
 \left|\psi_{n{\bf K}}^{\uparrow }  ({\bf r})\right|^2
-\left|\psi_{n{\bf K}}^{\downarrow }({\bf r})\right|^2 \right)d^2{\bf r}
\label{eqq3s}
\end{equation}
where the integral should be taken over the supercell with area $\Omega$, we expand the wavefunctions at ${\bf K}$ in (S1) in a plane wave basis as $\psi_{n{\bf K}}({\bf r}) =\sum_{\sigma {\bf G}} C_{n{\bf K}}^{\sigma {\bf G}}e^{i({\bf K}+{\bf G}).{\bf r}}$ and (S1) can be simplified to
\begin{equation}
\langle s_z \rangle_{n{\bf K}}=\sum_{\bf G}        \left(
 \left|C_{n{\bf K}}^{\uparrow   {\bf G}}\right|^2
-\left|C_{n{\bf K}}^{\downarrow {\bf G}}\right|^2  \right).
\label{eqq4s}
\end{equation}

\subsection{B. Berry curvature from phenomenological model}

The Berry curvature can be expressed in terms of the eigenvalues $\varepsilon_i({\bf q})$ and corresponding Bloch cell-periodic functions $u_i({\bf q})$ of the model Hamiltonian as 
\begin{equation}
B_i({\bf q})=2\mathfrak{I} \sum_{j\neq i}^4 
\frac{\langle u_i({\bf q})|\sigma_x|u_j({\bf q})\rangle 
      \langle u_j({\bf q})|\sigma_y|u_i({\bf q})\rangle }
      {(\varepsilon_i-\varepsilon_j)^2}.
\end{equation}
because the Hamiltonian is linear in ${\bf q}$. A quartic equation must be solved to determine the eigenvalues of the $4\times 4$ Hamiltonian matrix. The analytical solution is not practical for the general case and we instead solve it numerically. However, for $\lambda_R=\lambda_B=0$, we solve it analytically and use this to determine the phase space regions. $H$ is then ${\bf q}.{\bm \sigma} +\lambda _m\sigma _z+\lambda _{so}\sigma _zs_z$ and the energy eigenvalues are $\varepsilon _{4(2)}=\pm \sqrt{q^2+\left(\lambda _m-\lambda _{\text{so}}\right){}^2}$ and $\varepsilon _{3(1)}=\pm \sqrt{q^2+\left(\lambda _m+\lambda _{\text{so}}\right){}^2}$. The summation over occupied states of the Berry curvature is  
\begin{equation}
B({\bf q})=\frac{\lambda_m-\lambda_{\rm so}}{2\varepsilon_2^{3/2}}
          +\frac{\lambda_m+\lambda_{\rm so}}{2\varepsilon_1^{3/2}}
\end{equation}
The integral of the Berry curvature is 
\begin{equation}
\int_0^{\infty }B(q)qdq=\frac{1}{2}
\Big[\sgn(\lambda_m-\lambda_{\rm so})
    +\sgn(\lambda_m+\lambda_{\rm so})\Big]
\end{equation}
and therefore
\begin{eqnarray}
 \frac{1}{2\pi }\int B(q)d^2q=\left\{
\begin{array}{l}
 |\lambda_{\rm so}| \leq |\lambda_m| \rightarrow \sgn(\lambda_m)\equiv 1 \\
 \\
 |\lambda_m|        \leq |\lambda_{\rm so}| \rightarrow 0
\end{array}
\right.
\end{eqnarray}
and 
\begin{equation}
 \nu =\left\{
\begin{array}{l}
|\lambda_{\rm so}| \leq |\lambda_m|        \rightarrow 1+1\equiv 0(\text{NI}) \\
\\
|\lambda_m|        \leq |\lambda_{\rm so}| \rightarrow 0+1\equiv 1(\text{TI})
\end{array}
\right.
\end{equation}
As a result we can determine the $\mathbb{Z}_2$ invariant for four points in the phase space. Regions defined as all other points that can be connected to one of these four points without the gap closing will have the same invariant as that point. 

\subsection{C. Inversion Symmetric Structures}
\begin{figure}[t]
\includegraphics[width=8.0cm]{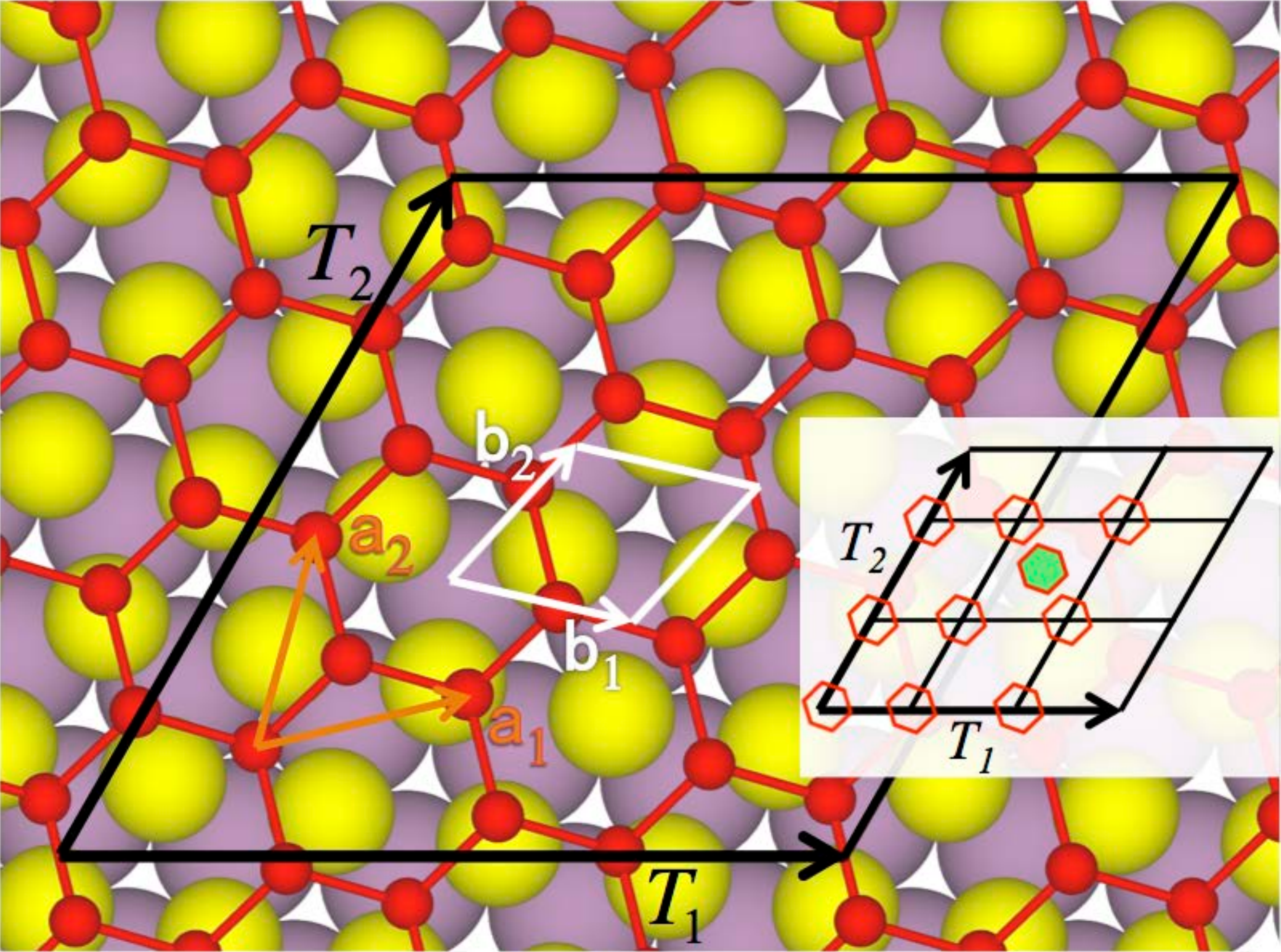}
\caption{Top view of a $\rm \overline{Ge}|$MoS$_2$ bilayer with three-fold rotation symmetry $C_3$ about the center of the unit cell that contains 26 Ge (red), 21 Mo (gray) and 42 S (yellow) atoms. The relative rotation angle is $24.8^{\circ}$. ${\bf a}_1$ and ${\bf a}_2$ (${\bf b}_1$ and ${\bf b}_2$) are the lattice vectors of $\rm \overline{Ge}$ (MoS$_2$). The supercell lattice vectors are $T_1=3{\bf a}_1-{\bf a}_2 = 4{\bf b}_1+{\bf b}_2$ and $T_2={\bf a}_1+3{\bf a}_2=-{\bf b}_1+4{\bf b}_2$. The central $\rm \overline{Ge}$ hexagon is shown schematically in the inset (green filled red hexagon); nine other configurations are sketched in which the central hexagon is displaced laterally breaking the $C_3$ symmetry. 
}
\label{SF1}
\end{figure}

For systems with inversion symmetry (IS), the topological invariant $\nu$ can be determined from the parities of the occupied states at the time reversal fixed points \cite{Fu:prb07s}. As a check of our phase space arguments for TI character, we therefore constructed IS systems and used both methods to study them. We began with an AS bilayer constructed so as to have threefold rotation symmetry by displacing the germanene in the $xy$ plane with respect to the MoS$_2$ ``substrate'' to make the threefold axes at the centre of the hexagonal rings (which are also inversion centres for the component layers) coincide. The IS trilayer was constructed by inverting the MoS$_2$ substrate layer through the germanene hexagonal ring inversion centre. These structures with threefold rotation and inversion symmetry were relaxed in two steps like we did for AS bilayers; first with respect to the height (h-IS) and then fully unconstrained (f-IS) in both cases maintaining the full symmetry. A top view of the AS $\rm \overline{Ge}|$MoS$_2$ bilayer supercell with $\theta = 24.8^{\circ}$ is shown in Fig.~\ref{SF1}. 

\subsection{D. Configuration Space Sampling}
For an AS $\rm \overline{Ge}|$MoS$_2$ bilayer, buckling brings one sublattice of germanene into closer contact with the substrate than the other leading to a non-vanishing mass term $\lambda_m$. Because it is not possible to study how the $\lambda$ parameters vary for general displacements and rotations of the germanene layer on the MoS$_2$ substrate, we studied their variation under (i) displacements and (ii) under rotations $\theta$ in the $xy$ plane parallel to the substrate separately.
 
\subsubsection{1. In-plane Displacements}
To study the effect of an in-plane displacement, we started with a height-optimized bilayer with $\theta = 24.8^{\circ}$ containing 89 atoms with threefold rotation symmetry, fixed the height of the germanene at its equilibrium separation and then displaced it by scanning the central Ge hexagon through the supercell as sketched in Fig.~\ref{SF1}. The result of sampling 10 points, given in Table~\ref{ST1}, is that the variation of the parameters is minimal. We interpret this in terms of the large unit cells and incommensurability of germanene and MoS$_2$. This means that the average environment seen collectively by the 13 Ge atoms of one sublattice does not vary as the layers are displaced parallel to one another. 

The $(6\alpha=3,6\beta=3)$ configuration with $C_3$ rotation symmetry can be studied with this symmetry enforced or not. The latter calculation requires three times as many k points and is three times more expensive. The small differences are a measure of the precision achievable with the chosen parameters. 

\begin{table}[t]
\caption{Dependence of the parameters extracted for an AS b-$\rm \overline{Ge}|$MoS$_2$ bilayer with $\theta=24.8^{\circ}$ on the $xy$ position of germanene with respect to MoS$_2$. The constant height of germanene above MoS$_2$ minimizes the energy for $\theta=24.8^{\circ}$. One Ge hexagon is placed at different positions in the supercell determined by ${\bf r}_{\rm hex}=\alpha {\bf T}_1+\beta {\bf T}_2$ where ${\bf T}_1$ and ${\bf T}_1$ are the supercell lattice vectors. Other parameters are as defined in the main article. The $(6\alpha=3,6\beta=3)$ configuration has $C_3$ rotation symmetry which is used explicitly to obtain the results in the top row. Not enforcing the symmetry yields slightly different results.}
\begin{ruledtabular}
\begin{tabular}{ccccccccccccc}
$6\alpha,6\beta$
       & $s$
              & $\Delta_{\bf K}$
                     & $\lambda_m$
                            & $\lambda_{\rm so}$ 
                                   	& $\lambda_{\rm so}^{\rm ind}$
                                            & $\lambda_R$ 
                                                   & $\lambda_B$ 
                                                           & $\nu$\\
\hline
$3,3$  & 0.83 & 5.55 & 7.95 & 11.60 & -1.29 & 5.72 & -0.56 & 1\\
\hline
$3,3$  & 0.81 & 6.38 & 7.49 & 11.60 & -1.29 & 5.72 & -0.56 & 1\\
$0,0$  & 0.81 & 6.38 & 7.49 & 11.60 & -1.29 & 5.74 & -0.56 & 1\\
$2,2$  & 0.81 & 6.38 & 7.49 & 11.60 & -1.29 & 5.74 & -0.56 & 1\\
$4,4$  & 0.81 & 6.38 & 7.49 & 11.60 & -1.29 & 5.74 & -0.56 & 1\\
$2,0$  & 0.81 & 6.37 & 7.50 & 11.60 & -1.29 & 5.74 & -0.56 & 1\\
$4,0$  & 0.81 & 6.37 & 7.50 & 11.60 & -1.29 & 5.74 & -0.56 & 1\\
$0,2$  & 0.81 & 6.37 & 7.50 & 11.60 & -1.29 & 5.74 & -0.56 & 1\\
$0,4$  & 0.81 & 6.37 & 7.50 & 11.60 & -1.29 & 5.74 & -0.56 & 1\\
$4,2$  & 0.81 & 6.37 & 7.50 & 11.60 & -1.29 & 5.74 & -0.56 & 1\\
$2,4$  & 0.81 & 6.37 & 7.50 & 11.60 & -1.29 & 5.74 & -0.56 & 1\\
\hline
av     & 0.81 & 6.37 & 7.50 & 11.60 & -1.29 & 5.74 & -0.56 & 1\\
\end{tabular}
\end{ruledtabular}
\label{ST1}
\end{table}
\vspace{-1em}

\begin{figure}[b]
\includegraphics[width=1.0\columnwidth]{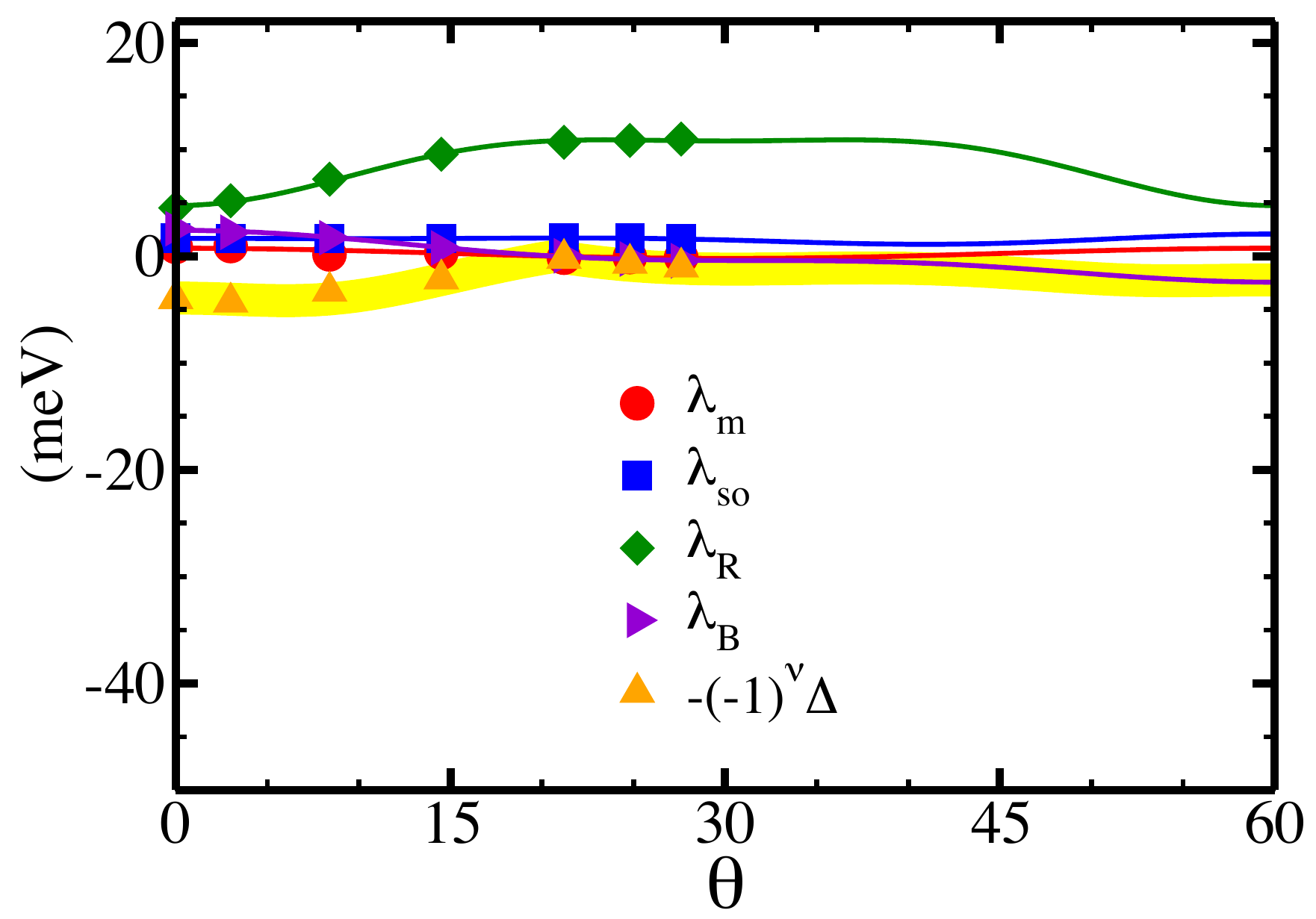}
\caption{$\lambda$ parameters as function of the angle $\theta$ for a fixed height of buckled germanene above MoS$_2$ that minimizes the energy for $\theta=24.8^{\circ}$ for b-$\rm \overline{Ge}|MoS_2$. The lines are fits to expressions with appropriate angle symmetries. $\Delta$ is the absolute size of the minimum gap, the thick yellow line and symbols are a signed gap where the topological invariant $\nu$ is 0 for a NI and 1 for a TI. }
\label{SF2}
\end{figure}

\subsubsection{2. In-plane Rotations}
\begin{figure}[b]
\includegraphics[width=1.0\columnwidth]{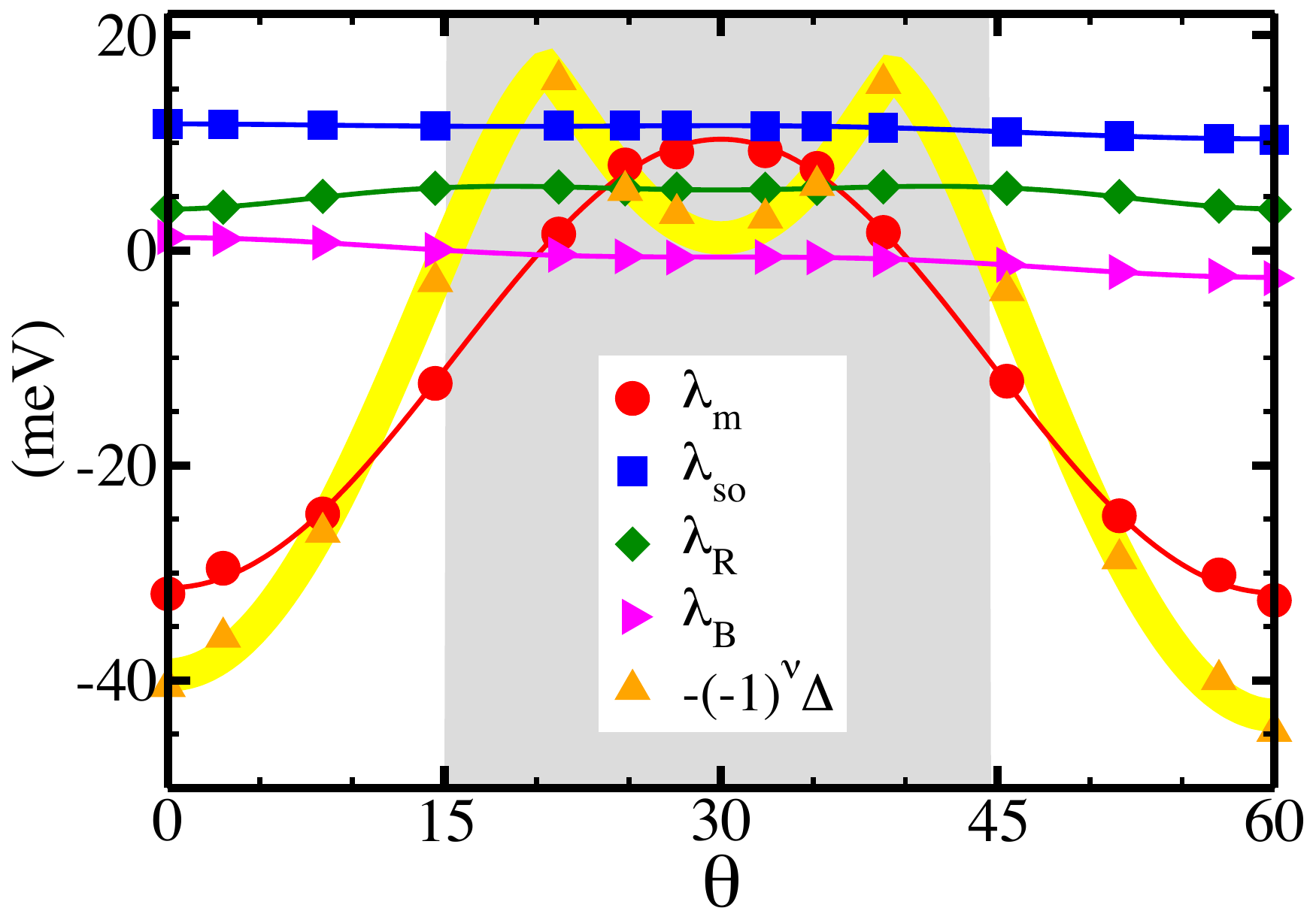}
\caption{$\lambda$ parameters as function of the angle $\theta$ for a fixed height of buckled germanene above MoS$_2$ that minimizes the energy for $\theta=24.8^{\circ}$ for b-$\rm \overline{Ge}|MoS_2$. The lines are fits to expressions with appropriate angle symmetries given in Eq.~(S8). $\Delta$ is the absolute size of the minimum gap, the thick yellow line and symbols are a signed gap where the topological invariant $\nu$ is 0 for a NI and 1 for a TI. The angles for which the system is TI is indicated by grey shading.  
} 
\label{SF3}
\end{figure}

Studying the effect of in-plane rotations of germanene with respect to the substrate is much more tedious and less systematic because the supercells required to model arbitrary rotations $\theta$ with an acceptable strain can be huge. For bilayers with threefold rotation symmetry (giving a factor 3 gain in computational expense because of the smaller irreducible BZ), we identify angles for which the supercell sizes are tractable in Tables \ref{ST2} and \ref{ST3}. Rotating germanene on MoS$_2$ has $2\pi/3$ periodicity. Because the individual layers have mirror symmetry, we only need to sample angles $\theta$ between 0 and $60^{\circ}$. For a maximum strain $\eta \leq 1\%$, we identify 14 angles between $0^{\circ}$ and $60^{\circ}$ with bilayer supercells containing a maximum of 341 atoms in Table~\ref{ST2} for p-$\rm \overline{Ge}|$MoS$_2$ and in Table~\ref{ST3} for b-$\rm \overline{Ge}|$MoS$_2$. The $\lambda$ parameters and gaps from these Tables are plotted in Fig.~\ref{SF2} on the same scale as used in Fig.~\ref{SF3}, for convenience repeated from the main article. For p-$\rm \overline{Ge}$, the $\theta$ dependence is weak but for the buckled case $\lambda_m$ changes quite dramatically. The angle dependence expected for the $\lambda$ parameters is
\begin{eqnarray}
\lambda_{m(R)}=&&a+b\cos{6\theta} +c(2\sin{6\theta} -\sin{12\theta})+d\cos{12\theta}, \nonumber \\
\lambda_B=&&a\cos{3\theta} +b(3\sin{3\theta} -\sin{9\theta})+c\cos 9\theta , \nonumber \\ 
\lambda_{\rm so}=&&a+b\cos{3\theta} +c(3\sin{3\theta} -\sin{9\theta}) +d\cos 6\theta + \nonumber \\
 &&e(2\sin{6\theta} -\sin{12\theta})+f\cos{9\theta} +g\cos{12\theta}.
\end{eqnarray}
\label{eqq8}
The coefficients obtained by fitting are given in Table \ref{ST4} for b-$\rm \overline{Ge}|$MoS$_2$. 

\subsection{E. General MoS$_2(\theta_1)|\bf \overline{Ge}|$MoS$_2(\theta_2)$ Trilayers}
If we now repeat our TB derivation of Eq.~1 for MoS$_2|\rm \overline{Ge}|$MoS$_2$ trilayers and neglect the direct interaction between ``substrate'' (S) and ``capping'' (C)  MoS$_2$ layers, then the contributions to the parameters in that equation describing the interaction with the neighbouring layers are additive and can be estimated from the AS $\rm \overline{Ge}|$MoS$_2$ bilayer calculations. For specific combinations of the angles $\theta_1$ and $\theta_2$, these estimates can be checked by performing explicit calculations for MoS$_2(\theta_1)|\rm \overline{Ge} |$MoS$_2(\theta_2)$ trilayers. The results of this confrontation of ``model'' predictions and first-principles \textsc{vasp} calculations are given in Table~\ref{ST5}. The agreement is excellent.
 
 The observation that the effect of the substrate (S) and capping (C) MoS$_2$ layers can be treated independently means that arbitrary S$|\rm \overline{Ge} |$C trilayers can be studied using the Hamiltonian  
\begin{equation}
H_{\bf K}({\bf q}) = \hbar v_F{\bf q}.{\bm \sigma} +\lambda_m \sigma_z + \frac{\lambda_R}{2}({\bm \sigma} \times {\bf s})_z + \lambda_{\rm so} \sigma_z s_z + \lambda _Bs_z  \nonumber
\label{eqq1s}
\end{equation}
where the parameters are determined independently for  S$|\rm \overline{Ge}$ and $\rm \overline{Ge} |$C bilayers. 
\vspace{1.5em}
\newpage
\begin{table*}[tb]
\caption{Hamiltonian parameters $\lambda_m$, $\lambda_{\rm so}$, $\lambda_{\rm so}^{\rm ind}$, $\lambda_R$ and $\lambda_B$ as a function of the rotation angle $\theta$ of the germanene layer about the symmetry axis in an AS p-$\rm \overline{Ge}|$MoS$_2$ bilayer with $C_3$ rotational symmetry. The height was optimized for $\theta=24.8^{\circ}$ and kept fixed for other angles.
$\eta$ is the substrate strain for supercells containing $N_{\rm \overline{Ge}}$ unit cells of germanene, $N_{\rm MoS_2}$ primitive unit cells of MoS$_2$ and $N_T$ atoms in total. 
The number $s$ characterizing the $z$ component of spin of two of the bands is defined in the main article.
$\Delta_{\bf K}$ (meV) is the band gap at the ${\bf K}$ point and $\nu$ is the topological invariant, 0 indicating a normal insulator and 1 indicating a non-trivial TI. 
}
\begin{ruledtabular}
\begin{tabular}{rccrcccrccrrc}
$\theta$
     & $N_{\overline{\text {Ge}}}$ 
          & $N_{\text{MoS}_2}$ 
               & $N_T$
                     & $\eta$
                           & $s$
                                  & $\Delta_{\bf K}$
                                         & $\lambda_m$
                                                 & $\lambda_{\rm so}$ 
                                                        & $\lambda_{\rm so}^{\rm ind}$
                                                                & $\lambda_R$ 
                                                                        & $\lambda_B$ 
                                                                                & $\nu$\\
\hline
27.6 & 19 & 31 & 131 & 0.2 & 0.01 & 0.18 & -0.09 & 1.61 & -0.50 & 10.97 &  0.00 &0\\
24.8 & 13 & 21 &  89 & 0.7 & 0.02 & 0.17 & -0.08 & 1.69 & -0.42 & 10.84 &  0.17 &0\\
21.2 & 49 & 79 & 335 & 0.8 & 0.06 & 0.65 & -0.14 & 1.70 & -0.41 & 10.67 &  0.46 &0\\
14.5 & 19 & 31 & 131 & 0.2 & 0.16 & 1.74 &  0.32 & 1.66 & -0.45 &  9.55 & 1.19 &0\\
 8.4 & 37 & 61 & 257 & 0.3 & 0.27 & 4.61 &  0.16 & 1.65 & -0.46 &  7.21 &  2.15 &0\\
 3.0 & 13 & 21 &  89 & 0.7 & 0.31 & 5.76 &  0.95 & 1.66 & -0.45 &  5.19 &  2.67 &0\\
 0.0 & 49 & 81 & 341 & 0.5 & 0.42 & 5.52 &  0.87 & 1.67 & -0.44 &  4.52 &  3.00 &0\\
\end{tabular}
\end{ruledtabular}
\label{ST2}
\end{table*}
\begin{table*}[b]
\caption{Hamiltonian parameters $\lambda_m$, $\lambda_{\rm so}$, $\lambda_{\rm so}^{\rm ind}$, $\lambda_R$ and $\lambda_B$ as a function of the rotation angle $\theta$ of the germanene layer of an AS b-$\rm \overline{Ge}|$MoS$_2$ bilayer with $C_3$ rotational symmetry about the symmetry axis. 
$\eta$ is the substrate strain for supercells containing $N_{\rm \overline{Ge}}$ unit cells of germanene, $N_{\rm MoS_2}$ primitive unit cells of MoS$_2$ and $N_T$ atoms in total. 
The number $s$ characterizing the $z$ component of spin of two of the bands is defined in the main article.
$\Delta_{\bf K}$ (meV) is the band gap at the ${\bf K}$ point and $\nu$ is the topological invariant, 0 indicating a normal insulator and 1 indicating a non-trivial TI.}
\begin{ruledtabular}
\begin{tabular}{rccrccrrccrrc}
$\theta$
     & $N_{\overline{\text {Ge}}}$ 
          & $N_{\text{MoS}_2}$ 
               & $N_T$
                     & $\eta$
                           & $s$
                                  & $\Delta_{\bf K}$
                                         & $\lambda_m$
                                                  & $\lambda_{\rm so}$ 
                                                          & $\lambda_{\rm so}^{\rm ind}$
                                                                  & $\lambda_R$ 
                                                                        & $\lambda_B$ 
                                                                                & $\nu$\\
\hline
60.0 & 49 &81 & 341 & 0.5 & 0.99 & 44.75 & -32.55 & 10.30 & -2.59 & 3.80 & -2.59 &0\\
57.0 & 13 &21 &  89 & 0.7 & 0.99 & 39.91 & -30.19 & 10.38 & -2.51 & 4.11 & -2.37 &0\\
51.6 & 37 &61 & 257 & 0.3 & 0.98 & 28.70 & -24.69 & 10.62 & -2.27 & 5.01 & -2.02 &0\\
45.5 & 19 &31 & 131 & 0.2 & 0.88 &  3.76 & -12.18 & 11.02 & -1.87 & 5.76 & -1.35 &0\\
38.8 & 49 &79 & 335 & 0.8 & 0.39 & 15.52 &   1.67 & 11.38 & -1.51 & 5.90 & -0.85 &1\\
35.2 & 13 &21 &  89 & 0.7 & 0.82 &  6.02 &   7.61 & 11.53 & -1.36 & 5.76 & -0.66 &1\\
32.4 & 19 &31 & 131 & 0.2 & 0.87 &  2.98 &   9.26 & 11.51 & -1.38 & 5.68 & -0.62 &1\\
27.6 & 19 &31 & 131 & 0.2 & 0.87 &  3.17 &   9.15 & 11.62 & -1.27 & 5.68 & -0.67 &1\\
24.8 & 13 &21 &  89 & 0.7 & 0.83 &  5.55 &   7.95 & 11.60 & -1.29 & 5.72 & -0.56 &1\\
21.2 & 49 &79 & 335 & 0.8 & 0.31 & 15.88 &   1.51 & 11.59 & -1.30 & 5.89 & -0.40 &1\\
14.5 & 19 &31 & 131 & 0.2 & 0.91 &  2.91 & -12.38 & 11.56 & -1.33 & 5.77 &  0.04 &0\\
 8.4 & 37 &61 & 257 & 0.3 & 0.98 & 26.22 & -24.52 & 11.65 & -1.24 & 5.02 &  0.72 &0\\
 3.0 & 13 &21 &  89 & 0.7 & 0.99 & 36.00 & -29.57 & 11.71 & -1.18 & 4.01 &  1.07 &0\\
 0.0 & 49 &81 & 341 & 0.5 & 0.99 & 40.58 & -31.95 & 11.77 & -1.12 & 3.81 &  1.23 &0\\
\end{tabular}
\end{ruledtabular}
\label{ST3}
\end{table*}

\begin{table}
\caption{The $\lambda$ parameters reported for an AS b-$\rm \overline{Ge}|$MoS$_2$ bilayer in Table~\ref{ST3} were fit to expressions with the angle dependence given in (S8) yielding the coefficients given here.}
\begin{ruledtabular}
\begin{tabular}{lrrrrrrr}
                             &  $a$  &   $b$  &  $c$  & $d$   & $e$   & $f$ & $g$   \\
\hline
$\lambda_m$                  &-10.86 & -20.98 &  0.00 &  0.01 &   -- &   -- &   --  \\
$\lambda_{\rm so}^{\rm ind}$ & -1.67 &   0.21 &  0.05 & -0.16 & 0.15 & 0.56 & -0.01 \\
$\lambda_R$                  &  5.30 &  -0.84 & -0.01 & -0.56 &   -- &   -- &   --  \\
$\lambda_B$                  &  0.00 &   0.47 &  0.03 &  0.44 &   -- &   -- &   --  \\
\end{tabular}
\end{ruledtabular}
\label{ST4}
\end{table}

\begin{table*}[b]
\caption{Hamiltonian parameters $\lambda_m$, $\lambda_{\rm so}$, $\lambda_{\rm so}^{\rm ind}$, $\lambda_R$ and $\lambda_B$ estimated (``model'') for MoS$_2(\theta_1)|\rm \overline{Ge} |$MoS$_2(\theta_2)$ trilayers with $C_3$ rotational symmetry using the results obtained from ab-initio calculations for AS b-$\rm \overline{Ge} |$MoS$_2(\theta)$ trilayers for a number of rotation angles $\theta_1$ and $\theta_2$  of ``substrate'' and ``capping'' MoS$_2$ layers that yield tractable supercells. 
``\textsc{vasp}'' indicates results obtained from explicit first principles calculations for the same trilayers.
$\eta$ is the substrate strain for supercells containing $N_{\rm \overline{Ge}}$ unit cells of germanene, $N_{\rm MoS_2}$ primitive unit cells of MoS$_2$ for each layer and $N_T$ atoms in total. 
The number $s$ characterizing the $z$ component of spin of two of the bands is defined in the main article.
$\Delta_{\bf K}$ (meV) is the band gap at the ${\bf K}$ point and $\nu$ is the topological invariant, 0 indicating a normal insulator and 1 indicating a non-trivial TI.}
\begin{ruledtabular}
\begin{tabular}{rrcccrccrrrccrc}
$\theta_1$ & $\theta_2$ &
                 & $N_{\overline{\text {Ge}}}$ 
                         & $N_{\text{MoS}_2}$ 
                         	& $N_T$
                               & $\eta$(\%)
                                    & $s$
                                         & $\Delta_{\bf K}$
                                                & $\lambda_m$
                                                         & $\lambda_{\rm so}$ 
                                                         		& $\lambda_{\rm so}^{\rm ind}$
                                                                 &$\lambda_R$ 
                                                                        & $\lambda_B$ 
                                                                                & $\nu$\\
\hline
$24.8^{\circ}$ & $24.8^{\circ}$  & model         & 13 & 21 & 152 & 0.7 & 1.00 & 20.62 & 0.00 & 10.31 & -2.58 & 0.00 &  0.00 & 1\\
$24.8^{\circ}$ & $24.8^{\circ}$  & \textsc{vasp} & 13 & 21 & 152 & 0.7 & 1.00 & 20.58 & 0.08 & 10.37 & -2.52 & 0.00 & -0.03 & 1\\
$3.0^{\circ}$  & $3.0^{\circ}$   & model         & 13 & 21 & 152 & 0.7 & 1.00 & 21.06 & 0.00 & 10.53 & -2.36 & 0.00 &  0.00 & 1\\
$3.0^{\circ}$  & $3.0^{\circ}$   & \textsc{vasp} & 13 & 21 & 152 & 0.7&1.00&21.16&-0.09&10.67&-2.22&0.00&0.06&1 \\
$3.0^{\circ}$  & $24.8^{\circ}$  & model         & 13 & 21 & 152 & 0.7 & 1.00 & 54.21 &37.52 & 10.42 & -2.47 & 1.71 &  1.63 & 0\\
$3.0^{\circ}$  & $24.8^{\circ}$  & \textsc{vasp} & 13 & 21 & 152 & 0.7 & 1.00 & 57.91 &38.39 &  9.47 & -3.42 & 2.40 &  1.52 & 0\\
\end{tabular}
\end{ruledtabular}
\label{ST5}
\end{table*}


%

\end{document}